\documentclass{an}
\usepackage{graphicx}
\usepackage{times}
\usepackage{fancyhdr}
\def\eg{{e.g.,~}}
\def\etal{{et\,al. }}
\def\asec{\ifmmode ^{\prime\prime}\else$^{\prime\prime}$\fi}

\def\degs{\ifmmode ^{\circ}\else$^{\circ}$\fi}
\def\amin{\ifmmode ^{\prime}\else$^{\prime}$\fi}
\def\asec{\ifmmode ^{\prime\prime}\else$^{\prime\prime}$\fi}
\def\farcs{\hbox{$.\!\!^{\prime\prime}$}}  

\def\cm{\mbox{\,cm}}

\def\cm3{\mbox{\,cm$^{-3}$}}
\def\kms{\mbox{\,km~s$^{-1}$}}

\def\kms{\mbox{\,km s$^{-1}$}}

\def\lsim{\!\!\!\phantom{\le}\smash{\buildrel{}\over
 {\lower2.5dd\hbox{$\buildrel{\lower2dd\hbox{$\displaystyle<$}}\over
                                 \sim$}}}\,\,}
\def\gsim{\!\!\!\phantom{\ge}\smash{\buildrel{}\over
{\lower2.5dd\hbox{$\buildrel{\lower2dd\hbox{$\displaystyle>$}}\over
                               \sim$}}}\,\,}

\def\Lya{{\rm\,Ly-$\alpha$~}}
\newcommand\HeII{He{\small II}~$\lambda$\,1640}
\newcommand\Hbeta{H$\beta$}
\newcommand\OII{[O{\small II}]~$\lambda$\,3727}
\newcommand\OIII{[O{\small III}]~$\lambda$\,5007}

\begin{document}

\title{Very high-resolution radio observations of HzRGs}

\author{M.~A. P\'erez-Torres\inst{1} \and
C. De Breuck\inst{2} \and W. van Breugel\inst{3} \and G. Miley\inst{4}
} 
\institute{
Instituto de Astrof\'{\i}sica de Andaluc\'{\i}a - CSIC, Apdo. Correos
3004, E-18080 Granada, Spain 
\and European Southern Observatory, Karl
Schwarzschild Stra\ss e 2, D-85748 Garching, Germany 
\and  Lawrence Livermore National Laboratory, PO Box 808, Livermore,
CA 94550, USA 
\and Leiden Observatory, University of Leiden, PO Box 9513, Leiden, 2300 RA, The Netherlands}

\date{Received; accepted; published online}

\abstract{We report on first results of an ongoing effort to image a
small sample of high-redshift radio galaxies (HzRGs) with
milliarcsecond (mas) resolution, using very-long-baseline
interferometry (VLBI) techniques.  
Here, we present 1.7 and 5.0~GHz VLBA observations of B3~J2330+3927, a
radio galaxy at $z$=3.087. Those observations, combined with 8.4~GHz
VLA-A observations, have helped us interpret the source radio
morphology, and most of our results have already been published
(P\'erez-Torres \& De Breuck 2005). In particular, we pinpointed the
core of the radio galaxy, and also detected both radio lobes, which
have a very asymmetric flux density ratio, $R>$11. Contrary to what is
seen in other radio galaxies, it is the radio lobe furthest from the
nucleus which is the brighest. Almost all of the Ly-alpha emission is
seen between the nucleus and the furthest radio lobe, which is also
unlike all other radio galaxies.  The values of radio lobe distance
ratio, and flux density ratio, as well as the fraction of core
emission make of B3~J2330+3927 an extremely asymmetric source, and
challenges unification models that explain the differences between
quasars and radio galaxies as due to orientation effects.
 }

\correspondence{torres@iaa.es}

\maketitle

\section{Introduction}

The combination of high-resolution radio observations with high 
angular resolution optical imaging and, more recently, 
with X-ray imaging, contributes important information to the 
study of high-redsifht radio galaxies (HzRGs, $z>2$). 
For example, 
McCarthy \etal\ (1987) and  Chambers \etal\ (1987) have clearly 
illustrated, by using radio, optical and UV
observations the existence of a radio-aligned 
optical line-emission and UV-continuum, strongly supporting 
the proposed scenario where 
a young radio AGN having powerful jets and ionizing
radiation can dramatically affect the environment of their forming
host galaxies.
This scenario has been confirmed by recent discoveries of
similarly radio-aligned X-ray emission (\eg\ Carilli \etal\ 2002,
Scharf \etal\ 2003). 

High-resolution radio data is also of high relevance in studying the
interaction of the radio sources with gas in the putative forming
galaxy (e.g., 100~kpc-scale \Lya\ halos). In particular, the
comparison of high-resolution radio and {\Lya}\ images may point to the
interaction between the propagating jet and the surrounding 
interstellar medium, and constrain the relative importance of
jet-induced star formation, shock ionized gas, and dust/electron
scattering in forming the observed optical galaxies.

\section{VLBI observations of HzRGs}

The enormous distances at which HzRGs lie, make it necessary to go
from 'high-resolution' to 'very high-resolution' radio observations,
if we are to better understand the physics of the individual objects
subject to study.  We have thus started a radio programme aimed at
studying with VLBI a small sample of high-redshift radio galaxies
whose peculiar radio morphologies suggest that strong interactions
with their surrounding medium are taking place, based on existing
high-resolution HST optical imaging.  Our strategy is to carry out
detailed morphological comparisons of radio and optical continua for
the sources of the sample, with the goal of distinguishing among the
effects of jet-induced star formation. VLBI images, of higher
resolution than HST optical images, can delineate the shocks
associated with the propagation of the jets, and even pinpoint
potential sites of star formation.  If there is no one-to-one
relationship between the radio and optical morphologies, this will
likely indicate that scattering is the dominant mechanism.  In
addition, a detailed morphological comparison with the \Lya\ maps may
be used to study the interaction between the
propagating jet and the surrounding primeval interstellar medium
(ISM). This ISM may cause the jets to bend and decollimate, and affect
the ionization and kinematics of the gas.

Our sample consists of four sources: B3~J2330+3927 ($z=3.1$),
B2~0902+34 ($z=3.4$), 4C~41.17 ($z=3.8$), and TN~J1338-1942 ($z=4.1$).
We observed B3~J2330+3927 on November-December 2004, using the VLBA at
1.7 and 5.0~GHz.  4C~41.17 and B2~0902+34 have been observed in June
2005 using simultaneously the European VLBI Network (EVN) and the
Multi-Element Radio-Linked Interferometer Network (MERLIN)
observations, and the data reduction is underway.  Here, we report
results from our observing campaign on B3~J2330+3927.  Throughout the
paper, we have assumed a $\Lambda$-dominated universe with
$\Omega_{\rm M}=0.3$, $\Omega_\Lambda = 0.7$, and $H_0 =
65\kms$~Mpc$^{-1}$.  At $z=3.087$, 1\arcsec\ corresponds to 8.2~kpc.

\section{B3~J2330+3927: A case study}

The radio galaxy B3~J2330+3927 has been studied at
several frequencies by De Breuck et al.~(2003), hereafter DB03.  The
8.4~GHz VLA observations of DB03 (see Fig.~\ref{fig,vla}) showed that
B3~J2330+3927 is a $\sim 2\asec$ wide source consisting of three radio
components, located in between two optical ($K-$band) objects ({\em a}
and {\em b}).  The triple radio morphology in the VLA A-array 8.4~GHz
map is reminiscent of other radio galaxies, where the host galaxy is
located at the position of the central radio component
Optical and near-IR Keck spectroscopy of
B3~J2330+3927 show AGN emission lines from object~{\em a}, while
object~{\em b} has a velocity offset of +1500~km/s with respect to
object~{\em a}. However, no emission lines are seen at the position of
the central radio source. Because the relative astrometric uncertainty
is $<$0\farcs4, the central radio component cannot be reconciled with
the AGN in object~{\em a}. De Breuck \etal\ (2003) proposed two
interpretations: (i) the northern radio component is the core,
implying a one-sided jet radio morphology, or (ii) the marginally
resolved central radio source is the radio core, and the AGN is
heavily obscured at this position. The latter explanation would be
inconsistent with the peak of the CO emission, which coincides with
object~{\em a}.  As we show below, only the higher angular resolution
provided by VLBI, together with spectral index information, has been
able to uniquely point to the correct AGN identification.

\begin{figure}
\resizebox{\hsize}{!}
{\includegraphics[width=8cm,angle=-90]{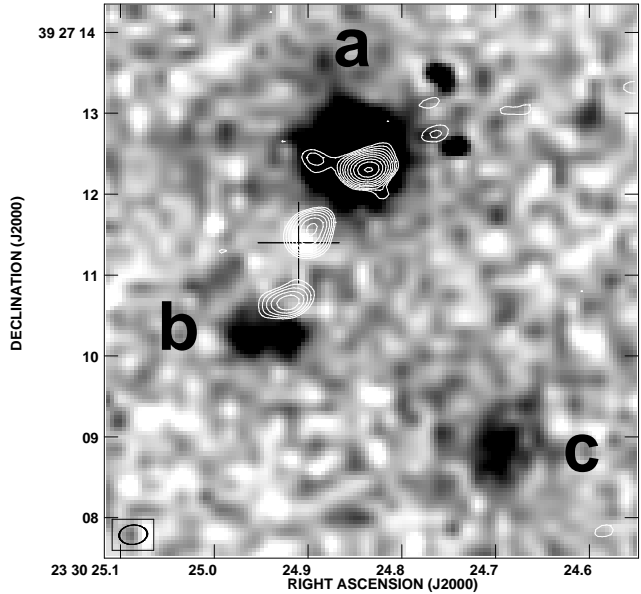}}
\caption{
8.4~GHz VLA A-array image of B3~J2330+3927 overlaid on a NIRC/Keck $K-$band image
(from De Breuck et al.~2003). The optical AGN
emission lines are located at the position of the bright object~{\em
a}, while the radio morphology suggests the core to be located in
between objects~{\em a} and {\em b}. The cross indicates the NVSS
position.
}
\label{fig,vla}
\end{figure}


We have recently published (P\'erez-Torres \& De Breuck 2005; 
hereafter PTDB)
new results on B3~J2330+3927, based on 
Very Long Baseline Array (VLBA) at 1.7 and 5~GHz carried out
on 2004 November 29 and December 9, respectively, as well
as on archival 1.4~GHz VLA data and 
our reanalysis of the 8.4~GHz VLA data of B3~J2330+3927 (DB03).
We analyzed the radio data using the Astronomical Image
Processing System ({\it AIPS}.
We used standard phase self-calibration techniques
within {\it AIPS} to obtain the images shown in Fig.~\ref{fig,vlavlba}.
We give in the following section an account of some of the results
obtained in PTDB, and refer the reader to that paper for technical 
details, as well as for a thorough account of the results obtained 
from our VLBA observations.

\begin{figure*}
\resizebox{\hsize}{!}
{\includegraphics[]{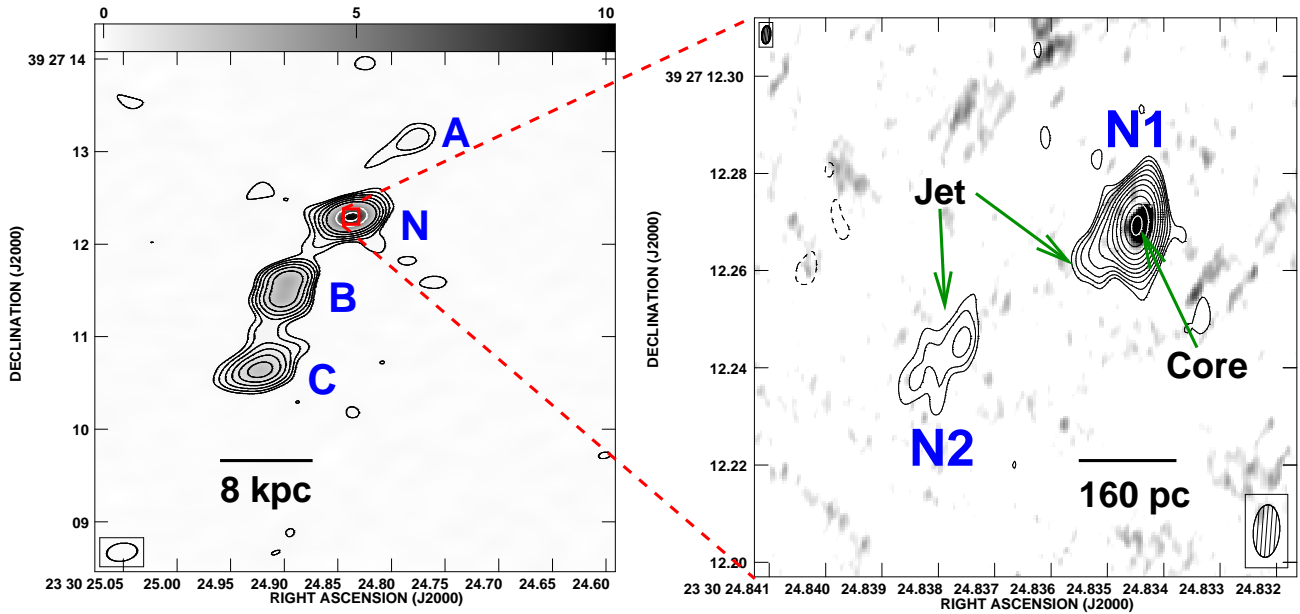}}
\caption{{\em Left:}8.4~GHz VLA A-array, uniformly weighted image of
B3~J2330+3927 on 30 March 2002.  The contour scheme is a geometric
progression in $\sqrt{3}$.  The first contour level is at 0.12\,mJy
beam$^{-1}$.  The synthesized beam is 0\farcs3$\times$0\farcs2 at
position angle -82\degs\.
The total cleaned flux
density in the image is $\approx$21.4~mJy and the image off-source rms
$\sim$40$\mu$Jy/beam (see left panel of Fig.~\ref{fig,vlavlba}).
Note the detection of component A, the faint, near radio lobe.  {\em
Right:} VLBA image of component N, as obtained from our VLBA
observations at 1.7~GHz (contours) and 5.0~GHz (greyscale).  Contour
levels are a geometric progression in $\sqrt{2}$, with the first
contour starting at 0.1\,mJy beam$^{-1}$. 
The off-source rms noise in the image was of $\sim 60\,\mu$Jy at both 1.7
and 5.0 GHz.
Note the protrusion
southeast of component N1, which corresponds to the jet emanating from
the radio core.}
\label{fig,vlavlba}
\end{figure*}

\section{Results}
\label{results}

Figure~\ref{fig,vlavlba} shows the reanalyzed 8.4~GHz VLA data on
B3~J2330+3927, with the archival 1.4~GHz data overlaid.  In addition
to the three components identified by DB03, we also detect a faint
component on the northwestern side of object~{\em a}.  In the
following, we call these components A, B, C, and N, as identified in
Fig.~\ref{fig,vla}. The newly found component,
A, shows for the first time the detection of the counter-jet in
B3~J2330+3927, providing support for the interpretation that the radio
component N, coincident with object~{\em a}, is the radio core.

We mapped the entire region, encompassing objects A through C, using
the VLBA.  Only components N and B were detected, both at 1.7 and
5.0~GHz. The left panel in Fig.~\ref{fig,vlavlba} shows the 8.4~GHz
map of these components, while the right panel displays a blow-up of
component N to show the VLBA data at their full resolution (for a
detailed report of those observations see PTDB).  Our 1.7~GHz VLBA data (contours) shows
that region N consists of a compact ($\lsim$4~mas in size) core-jet,
N1, and a jet feature, N2.  Our 5.0~GHz VLBA data only detected
component N1, which has a spectral index
$\alpha_{5.0}^{1.7}$=$-$0.2$\pm$0.1.  Component N2 was detected at
1.7~GHz, but not at 5.0~GHz, indicating that $\alpha_{5.0}^{1.7} \lsim
-$0.9.  Because component N1 has the flattest spectral index of all
components in our VLBA images, we identified it as the long-sought
radio core of B3~J2330+3927.

The identification of the core with component N1 confirms that C and the
newly found component, A, are the radio lobes of B3~J2330+3927.  Indeed,
wide-field (1\degr$\times$1\degr), deep 1.4~GHz VLA imaging 
(PTDB) excludes the presence of significant radio emission
beyond component A, and no other components are seen in the WENSS
or NVSS.  Thus, our
identification of component A with the outer radio lobe is robust.  
Now, from the
8.4~GHz reanalyzed data, we find that (a) the fraction of emission
from the nuclear component is $f_c$=0.50; (b) the ratio of the
core-lobe distances is $Q = (N1-C)/(N1-A) \approx$1.9; and (c) the ratio of
the flux densities of the further lobe to that of the closer lobe is
$R = S_{\rm C}/S_{\rm A}\gsim$11.

\begin{figure}
\resizebox{\hsize}{!}
{\includegraphics[angle=-90]{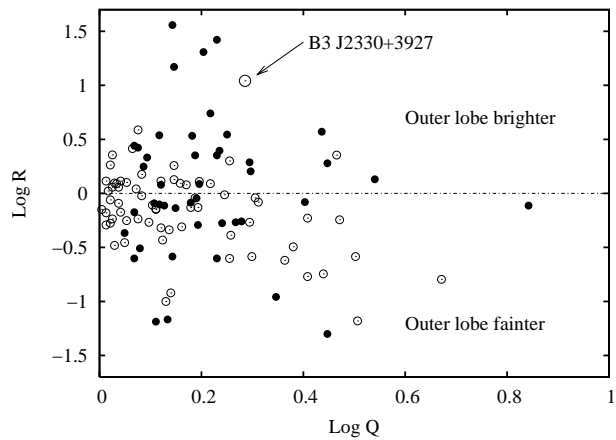}}
\caption{ The log $Q$ - log $R$ diagram for the radio galaxies (open
circles) and quasars (solid circles) of the Saikia et al. (2001)
sample, along with our case study source, B3~J2330+3927.  }
\label{fig,r-q}
\end{figure}

\section{Discussion}
\label{discussion}

The combined use of VLA and VLBA data, along with the existing
multiwavelength observations,  allows us to draw a 
consistent picture of the radio structure of B3~J2330+3927, 
revealing also several puzzling properties:
the compact, flat spectrum ($\alpha \sim -0.2$) component N,
is the radio core of B3~J2330+3927, and is located at the
position of the optical/near-IR type~II AGN and the peak of the
CO(4-3) emission.  The 8.4\,GHz VLA image reveals a previously
undetected, faint counterjet (A), confirmed by very deep, 1.4~GHz
VLA observations, and implying that A and C are the closer and farther
lobes, respectively, of the radio galaxy.  The radio source properties are extremely
asymmetric, as indicated by the values of $Q$, $R$, and $f_c$.

Best et al. (1995) have shown convincing evidence that quasars have more
asymmetric radio morphologies than radio galaxies, and this finding is
consistent with expectations of the unified model,
namely that quasars and radio galaxies are intrinsically similar, but
quasars are observed when the radio jet axis of the source is within
45\,\degr\ of the line of sight (Barthel 1989).  
We show in Fig.~\ref{fig,r-q} the log $Q$ - log $R$ diagram for the
78 radio galaxies and quasars from the 3CR and S4 samples selected by
Saikia et al. (1995, 2001). 
Note that there are no radio galaxies in the diagram with $\log R \gsim$0.6, 
in contrast with quasars. B3~J2330+3927, 
with a value of $\log R$=1.04, is thus a very bizarre radio galaxy, 
and looks more like a quasar.
Quasars also display larger $f_c$ values than radio galaxies (Saikia 1995).
Again, B3~J2330+3927, with half of its 8.4~GHz radio emission coming
from the core, looks more like a quasar.
Therefore, based on its radio morphology, B3~J2330+3927 would be
classified as a type~I AGN. However, the optical and near-IR Keck
spectra (Fig.~5 of DB03) clearly show only narrow emission lines with
relatively strong \HeII\ and very faint \Hbeta\ typical of type~II
AGN.  Moreover, the optical and near-IR continuum emission is weak and
relatively red, as is common in type~II AGN.

What is then the most likely explanation for the observed asymmetries
in the radio structure of B3~J2330+3927? 
Up to now, almost all cases of asymmetry in radio galaxies have been
explained by environmental effects. In the most detailed study to
date, McCarthy et al. (1991) find that for radio-loud type~II AGN, the
radio lobe closest to the core always lies on the same side of the
nucleus as the high surface brightness optical line emission. This is
contrary to what is seen in B3~J2330+3927. In fact, Fig.~4 of DB03 shows that
the \Lya emission extends from the radio core N to the southern radio
lobe~C, which is almost twice further away from the core than the faint
radio lobe~A. One could argue that the \Lya\ emission towards lobe~A
is quenched by dust, which is revealed by its strong far-IR emission
(Stevens et al. 2003, DB03). However, the spatial
profiles of the non-resonant \OII\ and \OIII\ lines in the Keck/NIRC
spectra of DB03 show a similar shape, with no emission towards lobe~A,
indicating that no substantial emission line flux is missing in the
\Lya\ profile. Environmental effects thus do not seem to play a
major role in the radio morphology of B3~J2330+3927.

If the observed asymmetries are due to relativistic beaming effects,
then the fact that B3~J2330+3927 is a clear type~II AGN restricts the
viewing angle with respect to the line of sight to $\phi$$>$45\degr
(Barthel 1989). However, the fraction of the emission from the nuclear
component, $f_c$=0.50 at 8.4~GHz, is very large for a radio galaxy,
and suggests the viewing angle must be then close to the
$\phi$=45\degr\ limit.  If this is case, the obtained arm-length ratio
$Q\approx$1.9 requires a jet velocity of $\beta \approx 0.45c$.  Now,
the observed ratio of the flux densities of components C and A is
$R\gsim$11. If components A and C are optically-thin, isotropically
emitting jets, their flux ratio should then be, within the
relativistic beaming scenario, $\lsim$6 if the jet is made of discrete
condensations, or $\lsim$4 if it is a continuous jet (an index of
$\alpha$=$-$0.9 has been assumed).  Our observed value of $R\gsim$11
is therefore difficult to explain within the standard relativistic
beaming framework.

The remaining explanation is an intrinsic difference in the radio jets
thus seems to be the most likely explanation of the observed
asymmetry. Examples of such radio galaxies are very rare in the
literature, and have only been reported at low redshifts. Sources
like B3~J2330+3927 and 4C~63.07 are difficult to reconcile with
predictions from the standard unified model.

It is also remarkable the lack of increased Ly-$\alpha$ emission 
near the region where the radio jet is
deflected between components N and B.  In fact, Fig.~2 of PTDB shows a
clear change in position angle between the jet-like feature emanating
from the core, N, and the line connecting components B1 and B2.
However, no increase in the Ly-$\alpha$ emission is seen in the 2D
spectrum (Fig.~4 of DB03), in contrast to similarly deflected radio
jets in other HzRGs, showing bright Ly-$\alpha$ emission at these
bendings (e.g., van Ojik et al.~1996).

\section{Summary}
\label{summary}

We have reported on 1.7 and 5.0~GHz VLBA observations of the
high-redshift radio galaxy B3~J2330+3927 ($z$=3.087).  The combination
of these observations with archival 8.4~GHz and 1.4~GHz VLA-A data has
helped us interpret the source radio morphology: we have pinpointed
the core and jet of the radio galaxy, and discovered a faint counter lobe,
which contrary to what has been seen in other radio galaxies lies much
farther from the nucleus than the brighter lobe.  The values of 
core-lobe distance ratio and flux density ratio, as well as the fraction
of core emission make of B3~J2330+3927 an extremely asymmetric source,
and challenges the standard unification model, which explains the
differences between quasars and radio galaxies as due to orientation
effects.

The results obtained from our VLBA observations of B3~J2330+3927
clearly show that very-long-baseline interferometry (VLBI)
observations of HzRGs can contribute crucial information towards a
detailed interpretation of the physics of those objects.

\acknowledgements 
We thank Robert Laing for stimulating discussions on
B3~J2330+3927, and D. Saikia for providing us with the values 
of $R$ and $Q$ used to produce the diagram in Fig.~\ref{fig,r-q}.   
This research has been partially funded by grant
AYA2001-2147-C02-01 of the Spanish Ministerio de Ciencia y
Tecnolog\'{\i}a.  
MAPT is supported by the programme Ram\'on y Cajal
of the Spanish Ministerio de Educaci\'on y Ciencia.  
The work by WvB at LLNL was performed under the auspices of the U.S.
Dept. of Energy and LLNL under
contract No. W-7405-Eng-48.
The National
Radio Astronomy Observatory is a facility of the National Science
Foundation operated under cooperative agreement by Associated
Universities, Inc.  This research has made use of NASA's Astrophysics
Data System.

\end{document}